\documentclass{aa}
\usepackage{times,amssymb}
\usepackage{graphicx}
\usepackage{amsmath}
\def \INT {{\it INTEGRAL}}
\begin{document}
   \title{\INT\ timing and localization performance}


   \author{R. Walter\inst{1,2}, P. Favre\inst{1,2}, P. Dubath\inst{1,2}, A. Domingo\inst{3}, G. Gienger\inst{4}, W. Hermsen\inst{5}, J. Pineiro\inst{4},
           \\ L. Kuiper\inst{5}, M. Schmidt\inst{4}, G. Skinner\inst{6}, M. Tuttlebee\inst{4}, G. Ziegler\inst{4}, T. J.-L. Courvoisier\inst{1,2}
             }
   \authorrunning{R. Walter, P. Favre et al.}
   
   \offprints{Roland.Walter@obs.unige.ch}

   \institute{\INT\ Science Data Centre, Chemin d'Ecogia 16, CH--1290 Versoix, Switzerland
         \and
             Observatoire de Gen\`eve, Chemin des Maillettes 51, CH--1290 Sauverny, Switzerland
         \and
            Laboratorio de Astrof\'{\i}sica Espacial y F\'{\i}sica Fundamental, POB 50727, E--28080 Madrid, Spain
         \and
             European Space Operation Center, Robert-Bosch-Str. 5, D-64293 Darmstadt, Germany
         \and
            SRON National Institute for Space Research, Sorbonnelaan 2, 3584 CA Utrecht, The Netherlands
         \and
            Centre d'Etude Spatiale des Rayonnements, 9 Avenue du Colonel Roche, F--31028 Toulouse Cedex 4, France  
             }

   \date{Received July 14, 2003; accepted August 26, 2003}

   \abstract{ In this letter we report on the accuracy of the attitude, misalignment, orbit and time
     correlation which are used to perform scientific analyses of the \INT\ data. The boresight attitude during
     science pointings has an accuracy of 3 ${\rm arcsec}$. At the center of the field, the misalignments have been 
     calibrated leading to a location accuracy of 4 to 40 ${\rm arcsec}$ for the different instruments. 
     The spacecraft position is known within 10 meters. The relative timing between
     instruments could be reconstructed within 10 $\mu{\rm sec}$ and the absolute timing within 40
     $\mu{\rm sec}$.   }

   \maketitle
%

\section{Introduction}

The timing and localization accuracies obtained from \INT\ observations depend on both the instrument 
performances (Ubertini et al, 2003; Vedrenne et al., 2003; Lund et al, 2003; Mas-Hesse et al., 2003) 
and on the quality of a set of auxiliary data. Here we present (1) the accuracy of those auxiliary 
data and (2) the overall performance figures. The accuracy of the star tracker boresight attitude is discussed
in Sect.~2, the instrument misalignment when compared to the boresight and the imaging performance in Sect.~3, the position 
of the spacecraft on its orbit in Sect.~4, and the absolute timing in Sect.~5. The measured accuracies 
meet the requirements that were defined prior to launch.

\section{Star tracker boresight attitude}

The \INT\ Attitude and Orbit Control Subsystem provides the necessary functionality to control the
spacecraft orientation during slews and stable pointing phases, in the so-called Inertial Pointing
and Slew mode. In this mode, a CCD star tracker is used to control the pitch and yaw attitude of the
spacecraft while a fine sun sensor is used to control the roll attitude around the telescope
boresight. Data from the star tracker and the fine sun sensor are also used for on-ground attitude 
reconstruction. The \INT\ 
spacecraft boresight is defined by the optical axis of the operational star tracker, which in-turn
is defined by the CCD reference system of the star tracker. The X--axis is the optical axis of the
operational star tracker, the Z--axis points perpendicular to the solar panels towards the Sun.
There are two star trackers on \INT\ and since launch only star tracker A has been used.

The star tracker is able to track up to 5 stars continuously with a full accuracy (i.e. a sub-pixel
resolution of 0.245 ${\rm arcsec}$ compared to a pixel dimension of about 40 ${\rm arcsec}$) and to produce a map of
the stars in the field of view. The mapping mode is used after each slew and reaction wheel bias
when the attitude is not precisely known. During a pointing the first star encountered during the 
mapping is tracked continuously and used as the guide star.

The position, magnitude and status of the tracked star and the raw fine sun sensor outputs, are
processed to construct the measured unit vectors to the stars in the star tracker CCD coordinate
system. Note that the alignments of the fine sun sensor heads have been calibrated during the
commissioning phase with respect to the star tracker CCD reference system. These measured vectors
calculated twice per second along with the catalogue vectors (corrected for proper motion) are then
used to provide an optimal estimate of the attitude that is converted into right ascension and
declination of the X and Z axis with respect to the J2000 reference system.

During dithering operations (Courvoisier, Walter, Beckmann et al. 2003) the typical sequence of
attitude records available in the historic attitude file consists of (1) pointing records for each
dither point, (2) closed loop slew and (3) settling records between dithering pointings, and (4)
open loop slew records for slews longer than about 4 degrees.

The {\it pointing} records provide attitude information during a period of stable pointing. If the
difference with respect to the attitude at the start of the pointing is less than 2.5 ${\rm arcsec}$, there
will be only one entry in the attitude file. This record provides the best average attitude estimate
for the duration specified. However, whenever the attitude varies by more than 2.5 ${\rm arcsec}$ as is the
case during a reaction wheel bias manoeuvre, new records are created every 2 seconds. The accuracy
during stable pointings in nominal conditions is about 3 ${\rm arcsec}$ (1$\sigma$). During science
pointings (with a typical duration of 30 minutes), the maximum attitude deviation is less than 3
${\rm arcsec}$ in 75\% of the cases and less than 7 ${\rm arcsec}$ in any case for all star tracker axis.

The {\it closed loop slew} records, generated every 10 seconds, provide instantaneous attitude reconstruction 
based on the tracked guide star and the Sun. The accuracy of the
attitude is slightly degraded because of the slew rate (up to 90 ${\rm arcsec/sec}$). Typically, the
accuracy of the attitude record is better than 10 ${\rm arcsec}$ (2$\sigma$).

The {\it settling period} records provide attitude reconstruction with a time resolution of 10
seconds when small manoeuvres or oscillations occur at the beginning or end of a pointing. The
accuracy of the attitude during the settling periods depends on the nature of the settling. After
slews, the accuracy of the attitude record is better than 10 ${\rm arcsec}$ (2$\sigma$). During reaction
wheel biases, when the spacecraft angular rates can be quite high, the accuracy is worse.

The {\it open loop slew} records provide an instantaneous attitude, which is determined based on all
the mappings that are commanded during the slew. The accuracy of the attitude is slightly degraded
because of the high slew rate (up to 200 ${\rm arcsec/sec}$). Typically, the accuracy of the attitude record
at the mapping time is around 20 ${\rm arcsec}$ (2$\sigma$).

After a long open slew, the spacecraft attitude can be up to 1$\degr$ off the targeted value. However, corrective 
slews are performed resulting in a final attitude that is within 1 ${\rm arcsec}$ (1 $\sigma$) and in any 
case within 7 ${\rm arcsec}$ of the planned position.

\section{Instruments misalignment}

The instrument misalignment is defined with respect to the star tracker orientation for which the
attitude solution is derived. Analysis software provided by the Integral Science Data Center (ISDC,
Courvoisier, Walter, Beckmann et al. 2003) takes those misalignments into account to extract proper
source positions.

We used pointings of the Crab Nebula (revolution 39 to 45) and of Cygnus X--1 (revolution 12 to 21)
to determine the misalignment of the ISGRI, SPI and JEM--X 1 instruments. As the Crab Nebula is not
a point source for JEM--X, we used the pointings of Cygnus X--1 and GRS 1915+105 (revolution 57) for
JEM--X 2 (JEM--X 1 was only operated until revolution 45). For each instrument, the pointings were
selected when the instrument was in a mode providing full imaging information.

For the SPI instrument, different combinations of data were tested and the best results were
obtained when using Crab positions derived from analyses of large groups of pointings. The
misalignment of PICSIT is assumed to be identical to that of ISGRI. For JEM--X, data were selected
only when the attitude was stable within 2.5 ${\rm arcsec}$. The misalignment of the OMC was determined
independently by comparing observed star positions to the expected ones on the CCD.

The misalignment matrix represents a rotation of the instrument frame with respect to the star
tracker frame and does not depend on the attitude of the spacecraft. This rotation is performed
through an Euler matrix $M$ (e.g., Arfken 1985) such that $M\vec{str}=\vec{instr}$ where $\vec{str}$
and $\vec{instr}$ are the unity vector columns of a source direction in the star tracker respectively
instrument coordinate reference system (the orientation of the coordinate systems being the same).

For each instrument and every selected pointing, we calculated the position of the source of
interest assuming a unity misalignment matrix and searched for the 3 Euler angles which minimized
the length of all the vectors $\vec{mes}-S^TMS\vec{cat}$ for each pointing, where $\vec{mes}$ and
$\vec{cat}$ are the unity vector columns of measured position and catalog source position
respectively in the star tracker coordinate reference system, and $S^T$ is the transformation matrix
from the star tracker to the equatorial coordinate frame.

The misalignment of the instruments are similar in direction and amplitude and dominated by a
misalignment of 9 ${\rm arcmin}$ of the star tracker when compared to the spacecraft structure. The resulting
matrices M are given in the auxiliary file \texttt{\small aux/adp/ref/irot/inst\_misalign\_20030717.fits} and 
are used automatically by the analysis software.

\begin{figure}[h]
\centering
\includegraphics[bb=18 144 600 718,clip,width=8cm]{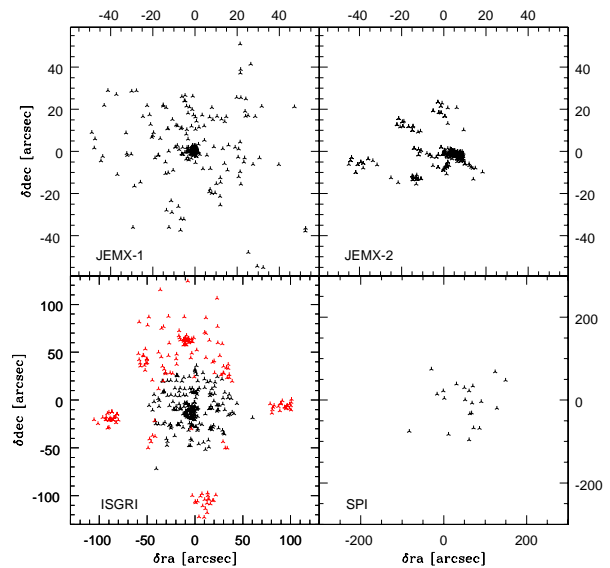}
\caption{Distribution of the deviations of the Crab Nebula (Cygnus X--1 for JEM--X2) positions as 
  derived with the misalignment matrices. For SPI, each point represents a group of 10 dither
  pointings. For ISGRI, the cluster of points close to the origin correspond to on-axis pointings,
  while the other clusters are related to the source position in the field of view during the
  pointings used in the analysis. Pointings with the Crab Nebula $>5\degr$ off axis are displayed in
  red. }
\label{residual}
\end{figure}

\begin{table}[h]
\caption{Residual error (radius) on the positions (at the center, on average, and at the border of
  the partially coded field of view) corrected for the misalignment. As the SPI analysis is based 
  on groups of pointings, only the mean value is given.}
\label{restab}
\begin{tabular}{lccc}
\hline\hline
Instrument & &Position residuals&   \\
                    & On-axis & Mean& Edge of the PCFOV \\
\hline
ISGRI  & 15\arcsec & 40\arcsec   & 140\arcsec\\ 
SPI    &           & 40\arcsec   & \\ 
JEM-X  &  4\arcsec & 18\arcsec   &  50\arcsec\\ 
\hline
\end{tabular}
\end{table}

Figure~\ref{residual} displays the distribution of the residual error position of the Crab Nebula
(Cygnus X--1 for JEM--X 2) peak flux, once the misalignment matrices are used in the analysis. The
residual error on the positions, listed in table~\ref{restab}, are related to the source off-axis
angle and increase from the center to the border of the partially coded field of view (PCFOV). More
information will be given in Favre (2004).

Analyzing the results obtained by the OMC on board centering algorithm, the accuracy of the OMC
misalignment matrix has been estimated to be better than half an OMC pixel ($\approx$ 9 ${\rm arcsec}$). No
significant differences are found at the border of the OMC field of view.

\section{Orbit data}

The spacecraft position is measured by a differential correction technique involving an orbital
model that takes into account the Earth potential, the gravitational effects of the Sun and of the
Moon, the effect of the Solar radiation pressure, and the momentum control manoeuvre data. This
model is constrained with ranging and Doppler measurement data providing line of sight position and velocity
that are regularly collected when the satellite is visible over the Redu ground station.  The
accuracy of the \INT\ orbit reconstruction is normally better than 10 meters along the line of
sight.

The orbital information is stored as a sequence of time ordered records. Each record is valid for
few hours on average and provides the undisturbed reference Keplerian Orbit and a polynomial
difference to this reference orbit to model the perturbations.  The reference frame for the orbital
information is the inertial mean geocentric equatorial system of J2000 (FK5). The system is centered
on the Earth gravitational center with the X axis pointing towards the mean vernal equinox. The X--Y
plane coincides with the mean Earth equatorial plane and the Z axis points towards the North.

\section{Timing information}

Measurements of the Crab pulsar indicate that absolute timing of \INT\ on-board events can be
derived with an accuracy of $\pm40$ $\mu{\rm sec}$. In this section we present some details of the
transformation of On Board Time (OBT) into Coordinated Universal Time (UTC), and we show that
at the beginning of the mission it has been affected by two kinds of unexpected delays. In addition
constant instrumental delays should also be taken into account. Those delay corrections are not 
integrated into the analysis software distributed by the ISDC. Delays provided in the tables 2 to 4
must therefore be corrected for by the users performing precise timing analysis.

\subsection{Time units and representations}

The \INT\ data analysis uses essentially three time systems:
\begin{enumerate}
\item The Earth Reception Time (ERT), expressed in coordinated Universal Time (UTC), is defined at
  the reception of every telemetry frame by the ground station. 
  The ERT is determined by atomic clocks located within each of the ground stations used by the
  mission. The ground stations are synchronized using the Global Positioning System. Unfortunately
  the synchronization was found to be affected by several problems (Sect.~5.2).
  
\item The Terrestrial Time (TT) is used to time tag, on-ground, products and physical events
  recorded within the instruments. The terrestrial time follows precisely the Atomic International
  Time (TAI) and does not suffer from leap seconds. In the data products terrestrial times are
  always formatted as double precision real in unit of \INT\ Julian Date (IJD), defined as the
  number of days since the 1st of January 2000 at 0h 0m 0s (TT) (IJD=JD--2451544.5).
  
\item The On Board Time (OBT) is defined by counting the number of pulses of an oscillator on board
  the spacecraft. All on board times are represented as 64 bit integers with a unit of $2^{-20}$ OBT
  second even if they are less precise in the telemetry. By convention, each time the on board clock
  is reset, the OBT is augmented by $2^{52}$ (this never happened at the time of writing). In FITS
  files, OBT are formatted as vectors of four 2-byte integers (for columns) or as 20 characters
  numeric string (for keywords).
\end{enumerate}

\subsection{Time correlation}

The time correlation is the relation between IJD and the on board time (OBT). It is derived from
measurements, in OBT unit, of the time at which specific telemetry frames leave the spacecraft
(more specifically the leading edge of the first frame bit). Those OBT measurements are then
correlated to the Earth Reception Time of the corresponding frames. Corrections for on board delays,
delays within each of the ground stations and light travel time are taken into account.

\begin{table}[b]
\caption[]{Offset to be added to the IJD derived by the time correlation.}
\label{bug}
{\footnotesize
\begin{tabular}{lcc}
\hline\hline
Period (UTC)   & Offset ($\mu{\rm sec}$)\\
\hline
before 2003-05-21 07:15&964 \\
from 2003-05-21 07:15 to 2003-07-03 14:23&867 \\
after 2003-07-03 14:23 &0 \\
\hline
\end{tabular}
}
\end{table}

The on board delays were calculated and calibrated on ground. Unfortunately, at the beginning of the
mission, the on board delay was taken into account with a wrong sign in the time correlation
software with the net effect that any IJD derived from an OBT should be corrected by a positive
offset as listed in table~\ref{bug}. This correction is not performed by the analysis software yet
and needs to be added to the terrestrial time derived from the time correlation.
On 2003-05-21 07:15 the telemetry rate of \INT\ was increased by 25\% resulting into a new correction offset. The software was corrected on 2003-07-03.

In addition, time offsets are observed at ground station handovers. This means that
systematic, and sometimes variable, delays are present between the timings of the different ground
station.  Besides the effect of the change of telemetry rate, those problems were dominated by a
miscalibration of the Goldstone DSS--24 station at the beginning of the mission, and by
synchronization problems within the Redu ground station. All calibrations were corrected on
2003-07-03.

Table~\ref{delay} lists the observed delays that should be added by the user to any IJD derived from 
an OBT to obtain an absolute time reference (this correction is not performed by the analysis software yet). 
The time derived from the VILSPA 2 station (except for the period
from 2003-05-21 to 2003-07-03) is taken as the reference. Note however that the UTC derived from
NASA ground stations leads the UTC derived from ESA ground stations by about 40 $\mu{\rm sec}$. For now
this should be considered as a residual systematic error.

\begin{table}[h]
\caption[]{Delays observed between the various ground stations and IJD (as defined by the VILSPA 2 
ground station). A positive delay indicates that for a given OBT, the IJD derived using the ground 
station in reference is smaller than it should be. A mean value and its uncertainty is provided for all periods during 
which the delays varied significantly. This means that values without reported errors are more accurate than
the others.}
\label{delay}
\begin{tabular}{llr}
\hline\hline
Ground Station       & Period& Delay\\
                                  &(UTC)              & ($\mu{\rm sec}$)\\
\hline
ESA Redu&before 2003-02-26 14:05&$-110\pm 10$\\
ESA Redu&2003-02-26 14:05 to 05-12 15:28&$-15\pm20$\\
ESA Redu&2003-05-12 15:28 to 05-15 15:25&$-105\pm10$\\
ESA Redu&2003-05-15 15:25 to 05-21 15:14&$0\pm5$\\
ESA Redu&2003-05-21 07:15 to 07-03 14:23&$74\pm5$\\
ESA VILSPA 2&before 2003-05-21 07:15&$0$\\
ESA VILSPA 2&2003-05-21 07:15 to 07-03 14:23&$73$\\
NASA DSS--16&before 2003-05-21 07:15&$42$\\
NASA DSS--16&2003-05-21 07:15 to 07-03 14:23&$34$\\
NASA DSS--24&before 2003-04-01 12:00&$-78081$\\
NASA DSS--24&2003-04-01 12:00 to 05-21 07:15&$54$\\
NASA DSS--24&2003-05-21 07:15 to 07-03 14:23&$43$\\
All&after 2003-07-03 14:23&$0$\\
\hline
\end{tabular}
\end{table}


\subsection{Instrument timing information}

A portion of the OBT counter is made available by the spacecraft to the \INT\ instruments which are
synchronized to the spacecraft at the beginning of every revolution (this synchronization involves a
delay of 15 $\mu{\rm sec}$ that is taken into account in table~\ref{idelay}). All instruments except the
spectrometer use the same clock pulse to time tag physical measurements. Those time tags (local on
board time) are transmitted in the telemetry, stored unchanged in the raw data, and converted to
full OBT unit in the subsequent steps of the data processing, providing a single OBT representation
for all times to the analysis software. The spectrometer camera uses its own internal oscillator to
tag events. Those time tags are converted to the standard OBT using specific time synchronization
information.

\begin{table}[t]
\caption[]{Delays between the event time and the event time tag (reconstructed for SPI) for the high
  energy instruments. The positive delay should be added to the event OBT time tag to get the absolute 
event time. The delay was calibrated on ground and measured in-flight comparing \INT\ and {\it RXTE}
Crab light curves in absolute phase.}
\label{idelay}
{\footnotesize
\begin{tabular}{lcc}
\hline\hline
Instrument       &Delay (ground calibration)&Delay (Crab timing)\\
                                  &($\mu{\rm sec}$)&($\mu{\rm sec}$) \\
\hline
SPI     &$134\pm10$ & $138\pm40$\\
IBIS    &$111\pm10$ & $135\pm40$\\
JEM--X   &$185\pm10$ & $216\pm40$\\
\hline
\end{tabular}
}
\end{table}

The delays between the actual event times and the instrument OBT time tags were measured on ground before the
launch (Alenia, 2002) and are given in the first column of table~\ref{idelay}. These delays are also
derived from flight data using contemporary \INT\ and {\it RXTE} observations of the Crab pulsar
(Kuiper, Hermsen, Walter et al. 2003).  For all \INT\ instruments and {\it RXTE}, the differences in
arrival times of the first (main) Crab peak in the pulse profile in radio and X-rays are measured.
The differences between the \INT\ and {\it RXTE} measurements, both using the Goldstone DSS--16
ground station, of the X-ray - radio delay is a measure of the instrumental delays, taking {\it
  RXTE} as the standard.  For {\it RXTE} an X-ray - radio delay of 268 $\pm$ 30 $\mu{\rm sec}$ was
determined, consistent with the 302 $\pm$ 67 $\mu{\rm sec}$ found by Rots, Jahoda, \& Lyne (2000)(see also Tennant,
Becker, Juda et al. 2001). The obtained differences are also listed in table 4. The quoted errors
reflect mainly the systematic uncertainty in the radio ephemeris due to variations in the dispersion
measure. The delays measured on ground and in-flight are fully consistent, and should be taken into
account for any absolute timing analysis. Formally those delays should be added to the on board times 
before applying the time correlation.

\begin{figure}[b]
\centering
\includegraphics[bb=14 14 499 490,clip,width=7.0cm]{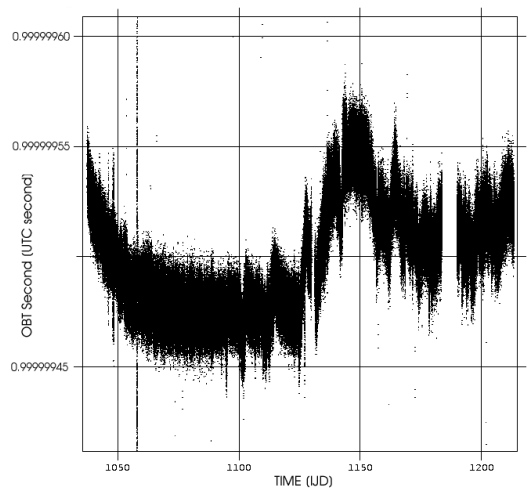}
\caption{OBT second corrected for thermal effects as a function of time.}
\label{clock}
\end{figure}

\subsection{On board clock stability}

The on board clock frequency varies with the oscillator temperature. Figure~\ref{clock} displays the
on board clock second (measured in Atomic International Time as measured by the ground station
atomic clocks), corrected for the thermal effects. The period of the on board clock is found
constant over a time scale of 6 months with an accuracy of 0.1 $\mu{\rm sec /sec}$. The residual slow
variations are not correlated to the orbital parameters nor with ground station usage, as expected.
This confirms that the light travel time is properly calculated.

\section{Conclusion}

The mean \INT\ boresight attitude during a science pointing is known to within 3 ${\rm arcsec}$. Attitude
instability during a pointing is limited to 7 ${\rm arcsec}$. The misalignment matrices provide on-axis
source localization accuracy for bright sources better than 4 ${\rm arcsec}$, 15 ${\rm arcsec}$ and 40 ${\rm arcsec}$ for
JEM--X, ISGRI and SPI respectively. This can be considered as a systematic uncertainty on the
localization accuracy. The localization is further limited for large off-axis angle and by
statistical uncertainties for faint sources.

The position of the spacecraft on the line of sight is known within $\pm$10 meters. Systematic
uncertainties on the relative timing have been calibrated within 10 $\mu{\rm sec}$. An additional
systematic uncertainty of about 40 $\mu$sec should be taken into account when comparing \INT\ timing
with absolute times.

Instrument misalignment matrices are integrated in the analysis software system provided by the ISDC
(Off-line Scientific Analysis version 2.0). This is not yet the case for the time delays given in
tables~\ref{bug} to ~\ref{idelay}. For accurate timing analysis users need to take those time 
delays into account themself.

\end{document}